\documentclass[11pt]{article}
	\usepackage{amsmath}
	\usepackage{amssymb}
	\usepackage{graphicx}

\begin{document}
	
	\setcounter{figure}{0}
	\setcounter{table}{0}
	\setcounter{footnote}{0}
	\setcounter{equation}{0}
	
	\noindent {\Large\bf POST-LINEAR METRIC OF A SOLAR SYSTEM BODY}
	\vspace*{0.7cm}

	\noindent\hspace*{1cm} S. Zschocke$^1$  \\[0.2cm]
	\noindent\hspace*{1cm} $^1$ Institute of Planetary Geodesy, TU Dresden - Germany - sven.zschocke@tu-dresden.de\\
	
	\vspace*{1cm}
	
	\noindent {\large\bf ABSTRACT.} 

	A precise modeling of light trajectories in the solar system on the sub-micro-arcsecond and nano-arcsecond scale of accuracy  
	requires the metric tensor of solar system bodies in post-linear approximation. The Multipolar Post-Minkowskian formalism  
	represents a framework for determining the metric density in the exterior of a compact source of matter, which can be regarded as massive  
	solar system body. The knowledge of the metric density, frequently been called gothic metric,  
	allows to deduce the metric tensor. Some aspects are considered about how to determine 
	the metric density and the metric tensor from the field equations of gravity.

	\vspace*{1cm}
	
	\noindent {\large\bf 1. INTRODUCTION}
	\smallskip
	
	An advancement in astrometric science towards sub-micro-arcsecond and nano-arcsecond level in angular measurements of celestial objects  
	requires considerable progress 
	in the theory of light propagation through the curvilinear space-time of the solar system. In curved space-time the  
	light signals propagate along null-geodesics, governed by the geodesic equation which reads  
	$\ddot{x}^{\alpha}\!\left(\lambda\right) + \Gamma^{\alpha}_{\mu\nu}\;\dot{x}^{\mu}\!\left(\lambda\right)\,\dot{x}^{\nu}\!\left(\lambda\right) = 0$,  
        where $x^{\alpha}\!\left(\lambda\right)$ is the four-coordinate of the light signal as function of the affine curve parameter $\lambda$, 
	a dot means total derivative with respect to $\lambda$, and the Christoffel symbols  
	$\Gamma^{\alpha}_{\mu\nu} = g^{\alpha\beta}\left(g_{\beta\mu\,,\,\nu} + g_{\beta\nu\,,\,\mu} - g_{\mu\nu\,,\,\beta}\right)/2\;$ are functions of the  
	metric tensor $g_{\alpha\beta}$, and a comma denotes partial derivative with respect to the four-coordinates, e.g.  
	$f_{\,,\,\mu} = \partial f /\partial x^{\mu}$ and $f_{\,,\,\mu\nu} = \partial^2 f /\partial x^{\mu}\,\partial x^{\nu}$, etc.  
	Accordingly, a precise modeling of light trajectories implies a precise knowledge of the metric of solar system bodies.   
	The metric tensor can be series expanded in powers of the gravitational constant $G$, called post-Minkowskian expansion,  
	\begin{equation}
	g_{\alpha\beta}\left(x\right) = \eta_{\alpha\beta} + \sum\limits_{n=1}^{\infty} G^n\,h^{\left({\rm nPM}\right)}_{\alpha\beta}\left(x\right)  
	\label{post-Minkowskian_expansion_1} 	
	\end{equation} 

	\noindent
        where the first and second term, $h^{\left({\rm 1PM}\right)}_{\alpha\beta}$ and $h^{\left({\rm 2PM}\right)}_{\alpha\beta}$,  
	are the linear and post-linear term of the metric perturbation, which are required for determining the light trajectory on the  
	sub-micro-arcsecond and nano-arcsecond scale of accuracy. The orthogonality relation $g^{\alpha\rho}\;g_{\rho\beta} = \delta^{\alpha}_{\beta}$ enables  
        to switch between the contravariant and covariant components of the metric tensor.  

	The Multipolar Post-Minkowskian (MPM) formalism represents a perturbative approach for  
	determining the metric density, $\overline{g}^{\alpha\beta}$,  
	in the exterior of a compact source of matter, defined by 
        \begin{equation}
		\overline{g}^{\alpha\beta} = \sqrt{- g\;}\;g^{\alpha\beta} \quad \quad {\rm or} \quad\quad  
		g^{\alpha\beta} = \sqrt{- \overline{g}\;}\;\,\overline{g}^{\alpha\beta}  
        \label{Metric_Density}
        \end{equation}

        \noindent 
	where $g = {\rm det}\left(g_{\rho\sigma}\right)$ and $\overline{g} = {\rm det}\left(\overline{g}_{\rho\sigma}\right)$ is the 
	determinant of the covariant components of the metric tensor and metric density, respectively. 
	The post-Minkowskian expansion of the metric density reads   
        \begin{equation}
	\overline{g}^{\alpha\beta}\left(x\right) = \eta^{\alpha\beta} 
	- \sum\limits_{n=1}^{\infty} G^n\,\overline{h}_{\left({\rm nPM}\right)}^{\alpha\beta}\left(x\right)  
        \label{post-Minkowskian_expansion_2}   
        \end{equation}

        \noindent
	where the first and second term, $\overline{h}_{\left({\rm 1PM}\right)}^{\alpha\beta}$ and $\overline{h}_{\left({\rm 2PM}\right)}^{\alpha\beta}$,  
        are the linear and post-linear term of the gothic metric perturbation.  
        The orthogonality relation $\overline{g}^{\alpha\rho}\;\overline{g}_{\rho\beta} = \delta^{\alpha}_{\beta}$ enables to switch between the contravariant 
	and covariant components of the gothic metric.  

	The MPM formalism determines the metric density in the exterior of a massive body, having arbitrary shape, inner structure, oscillations, 
	and rotational motions. Due to Eq.~(\ref{Metric_Density}) the know\-ledge of the metric density allows to deduce the metric tensor.    
	In what follows, some aspects are considered about how to obtain the metric density and metric tensor from the field equations.

	\vspace*{0.7cm}
	\noindent {\large\bf 2. THE FIELD EQUATIONS OF GRAVITY}
	\smallskip
	
	The field equations relate the metric tensor $g_{\mu\nu}$ to the stress-energy tensor of matter $T_{\mu\nu}\,$,  
        \begin{eqnarray}
        R_{\mu\nu} - \frac{1}{2}\,g_{\mu\nu}\,R = \frac{8\,\pi\,G}{c^4}\,T_{\mu\nu}  
        \label{exact_field_equations}
        \end{eqnarray}

\noindent
where $R_{\mu\nu} = \Gamma^{\rho}_{\mu\nu\,,\,\rho} - \Gamma^{\rho}_{\mu\rho\,,\,\nu}  
+ \Gamma^{\rho}_{\sigma\rho}\,\Gamma^{\sigma}_{\mu\nu} - \Gamma^{\rho}_{\sigma\nu}\,\Gamma^{\sigma}_{\mu\rho}\,$ is the Ricci tensor 
and $R = g^{\mu\nu}\,R_{\mu\nu}$ is the Ricci scalar.  
The field equations constitute a set of ten coup\-led non-linear partial differential equations for
the ten components of the metric tensor $g_{\mu\nu}$ of space-time, which in differential geometry is modeled by a   
semi-Riemannian manifold ${\cal M}$.  
The contracted Bianchi identities imply that  
only six of these field equations (\ref{exact_field_equations}) are independent, which
determine the ten components of the metric tensor up to a passive coordinate transformation
(keep points of manifold fixed and change coordinates) from the old $\{y\}$ to the new coordinate system $\{y^{\prime}\}$,
\begin{equation}
y^{\mu} \rightarrow y^{\prime\,\mu}\,.    
\label{Coordinate_Transformation_1} 
\end{equation}

\noindent
The field equations (\ref{exact_field_equations}) are invariant under these (infinitely many) coordinate transformations, 
known as passive general covariance of the field equations.  
That means, if the set $({\cal M},  {\bf g})$ is a solution of the field equations, then the set  
$({\cal M}, {\bf g}^{\prime})$ is also a solution of the same field equations, where 
$\displaystyle g^{\prime}_{\alpha\beta} = A^{\mu}_{\alpha}\,A^{\nu}_{\beta}\,g_{\mu\nu}$ is   
the metric tensor in these new coordinates with $A^{\mu}_{\alpha}$ being the Jacobian matrix  
$A^{\mu}_{\alpha} = \partial y^{\mu}/\partial y^{\prime \alpha}$ of the passive coordinate transformation. 
These sets are physically equivalent and describe the same physical system. The metric tensors  
have different components in different coordinate systems, $g^{\prime}_{\alpha\beta} \neq g_{\mu\nu}$, 
but as geometrical objects they are equal, ${\bf g}^{\prime} = {\bf g}$, because  
they attribute the same distance to the same pair of points ${\cal P}$ and ${\cal Q}$ of the manifold: 
$d_{{\bf g}^{\prime}}\left({\cal P},{\cal Q}\right) = d_{{\bf g}}\left({\cal P},{\cal Q}\right)$ (infinitesimal distance of these pairs is assumed).  
For later purposes it is useful to consider an active coordinate transformation (keep coordinates fixed and change points of manifold), 
\begin{equation}
\Psi: {\cal M} \rightarrow {\cal M} 
\label{Diffeomorphism_M_M} 
\end{equation}

\noindent 
which is a $C^{\infty}$ differentiable mapping of each point of the mani\-fold reversibly unique to another image point of the same manifold,  
${\cal P} \rightarrow \Psi\left({\cal P}\right)$. Hence, the coordinates are changed  
$y^{\mu}\left({\cal P}\right) \rightarrow y^{\prime\,\mu}\left({\cal P}\right)$.  
The field equations (\ref{exact_field_equations}) are invariant under these (infinitely many) diffeomorphisms, known as active general covariance  
of the field equations. That means,  
if the set $\left({\cal M}, {\bf g}\right)$ is a solution of the field equations, then the set 
$\left({\cal M}, {\bf g}^{\prime}\right)$ is also a solution of the 
same field equations, where ${\bf g}^{\prime} = \Psi^{\ast} {\bf g}$ is the pullback of the metric tensor,  
$g^{\prime}_{\alpha\beta} = A^{\mu}_{\alpha}\,A^{\nu}_{\beta}\,g_{\mu\nu}$,   
with $A^{\mu}_{\alpha}$ being the Jacobian matrix  
$A^{\mu}_{\alpha} = \partial y^{\mu}/\partial y^{\prime \alpha}$ of the active coordinate transformation. 
These sets are physically equivalent and describe the same physical system (Section $7.1$ in {\it Hawking,Ellis} (1974);  
for the associated problem of Leibniz Equivalence see {\it Earman,Norton} (1987) and {\it Lusanna,Pauri} (2006)). 
These metric tensors attribute the same distance of a pair of points of the manifold and their images,  
$d_{{\bf g}^{\prime}}\left({\cal P},{\cal Q}\right) = d_{{\bf g}}\left(\Psi\left({\cal P}\right),\Psi\left({\cal Q}\right)\right)$ 
(infinitesimal distance of these pairs and their images is assumed). 
But these metric tensors are not equal, ${\bf g}^{\prime} \neq {\bf g}$, because they attribute different distances to the same pair of points of the manifold:  
$d_{{\bf g}^{\prime}}\left({\cal P},{\cal Q}\right) \neq d_{{\bf g}}\left({\cal P},{\cal Q}\right)$ (e.g. {\it Gaul, Rovelli}, (2000)). 
However, if a Killing vector field exists  
on ${\cal M}$ and the diffeomorphism $\Psi$ proceeds along the congruence of that Killing vector field, then the metric and pullback metric  
are equal, ${\bf g}^{\prime} = {\bf g}$, and the diffeomorphism is an isometry (Section $2.6$ in {\it Hawking,Ellis} (1974)).  

\newpage

        \noindent {\large\bf 3. LANDAU-LIFSCHITZ FORMULATION OF GRAVITY}
        \smallskip

The theory of gravity has a geometrical interpretation in physical curvilinear space-time and a field-theoretical interpretation  
in auxiliary flat space-time (e.g. text below Eq.~(11) in {\it Gupta} (1954) or Section $8.4$ in {\it Feynman} (1995) or part $5$ in Box $17.2$  
in {\it Misner,Thorne,Wheeler} (1973)); for an excellent historical overview we refer to {\it Brian Pitts, Schieve} (2018).  
So one distinguishes between a physical manifold ${\cal M}$ covered by curvilinear coordinates  
$y^{\mu}$ and endowed with metric $g_{\mu\nu}\left(y\right)$, a flat background manifold ${\cal M}_0$ covered  
by curvilinear coordinates $x^{\alpha}$ and endowed with metric $g^0_{\alpha\beta}\left(x\right)$, and a diffeomorphism  
\begin{equation}
\Phi: {\cal M}_0 \rightarrow {\cal M}  
\label{Diffeomorphism_M_M0} 
\end{equation}

\noindent
which is a $C^{\infty}$ differentiable mapping of each point $q \in {\cal M}_0$ of the flat background manifold ${\cal M}_0$ reversibly unique to another point  
$p \in {\cal M}$ of the physical manifold ${\cal M}$ (hence ${\rm dim}\,{\cal M}_0 = {\rm dim}\,{\cal M}$); it is not relevant whether  
(\ref{Diffeomorphism_M_M0}) exists everywhere or only on finite domains  
$\Phi: {\cal V} \subseteq {\cal M}_0 \rightarrow {\cal U} \subseteq {\cal M}$. 

The field equations (\ref{exact_field_equations}) are not invariant under (\ref{Diffeomorphism_M_M0}), because the manifolds ${\cal M}$ and ${\cal M}_0$ 
are different with respect to their geometrical properties: 
the curvature tensor of ${\cal M}$ expressed in terms of $g_{\mu\nu}\left(y\right)$ is non-zero, $R^{\mu}_{\alpha\nu\beta}\left(y\right) \neq 0$,
in any coordinate system $\{y\}$ which maps the physical manifold,  
while the curvature tensor of ${\cal M}_0$ expressed in terms of $g^0_{\alpha\beta}\left(x\right)$  
vanishes, $R^{\mu}_{\alpha\nu\beta}\left(x\right) = 0$, in any coordinate system $\{x\}$ which maps the flat background manifold. In particular, the metric tensor  
${\bf g}^0$ of ${\cal M}_0$ (e.g. in Cartesian coordinates ${\bf g}^0$ is given by $\eta_{\alpha\beta} = {\rm diag}\left(-1,+1,+1,+1\right)$) and 
the metric tensor ${\bf g}$ of ${\cal M}$ can never be related by a pullback: ${\bf g}^0 \neq \Phi^{\ast} {\bf g}$.  

But the diffeomorphism (\ref{Diffeomorphism_M_M0}) is an active coordinate transformation, which makes it possible  
to pullback the metric tensor ${\bf g}$ of the physical manifold ${\cal M}$ (given by $g_{\mu\nu}\left(y\right)$) to the metric tensor 
$\Phi^{\ast} {\bf g}$ which propagates as tensorial field on the flat background ${\cal M}_0\,$ 
(given by $g_{\alpha\beta}\left(x\right)$) 
\begin{equation}
	g_{\alpha\beta}\left(x\right) = \frac{\partial y^{\mu}}{\partial x^{\alpha}}\, \frac{\partial y^{\nu}}{\partial x^{\beta}}\, g_{\mu\nu}\left(y\right)\,.  
\label{Diffeomorphism_M_M0_metric_1} 
\end{equation}

\noindent
In the same way, the Ricci tensor and energy-momentum tensor on ${\cal M}$ are pulled back on ${\cal M}_0\,$. By means of these relations the 
field equations of gravity (\ref{exact_field_equations}) on the physical manifold ${\cal M}$ can be pulled back to field equations on the flat  
background manifold ${\cal M}_0$. Then, the sets $\left({\cal M}, {\bf g}\right)$ and $\left({\cal M}_0, \Phi^{\ast}\,{\bf g}\right)$ are physically equivalent,  
iff the metric tensor ${\bf g}$ on the physical manifold ${\cal M}$ is determined by the field equations (\ref{exact_field_equations}),  
while the pulled-back metric tensor $\Phi^{\ast}\,{\bf g}$ on the flat background manifold ${\cal M}_0$ 
(i.e. $g_{\alpha\beta} = \Phi^{\ast\,\mu\nu}_{\alpha\beta}\,g_{\mu\nu}$ in Eq.~(\ref{Diffeomorphism_M_M0_metric_1})) is determined by the pulled-back 
field equations on ${\cal M}_0$ (cf. Section $7$ in {\it Hawking,Ellis} (1974), especially text below Eq.~(7.51) in {\it Hawking,Ellis} (1974), as well  
as text below Eq.~(7.10) in {\it Carroll} (2013)).  

In the Landau-Lifschitz formulation one makes a detour and does not consider the metric tensor $g_{\mu\nu}\left(y\right)$ but the metric density  
$\overline{g}^{\mu\nu}\left(y\right)$, which is pulled back from the physical manifold to the flat background manifold. 
A detailed mathematical representation of the Landau-Lifschitz formulation is given  
by Sections $1$ and $2$ in {\it Petrov,Kopeikin,Lompay,Tekin} (2017) as well as by Section $7$ in {\it Hawking,Ellis} (1974).  
These field equations take the following form (cf. Eqs.~(20.20) - (20.22) in {\it Misner,Thorne,Wheeler} (1973), Eq.~(6.6) in {\it Poisson,Will} (2014)), 
\begin{equation}
	H^{\alpha\rho\beta\sigma}_{\;\;\;\;\;\;\;\;\;,\,\rho\sigma}\left(x\right) 
	= \frac{16\,\pi\,G}{c^4}\left(-g\left(x\right)\right) \left(T^{\alpha\beta}\left(x\right) + t_{\rm LL}^{\alpha\beta}\left(x\right)\right). 
\label{Landau_Lifschitz_1}
\end{equation}

\noindent
The l.h.s. is the Landau-Lifschitz superpotential,  
$H^{\alpha\rho\beta\sigma} = \overline{g}^{\alpha\beta}\,\overline{g}^{\rho\sigma} - \overline{g}^{\alpha\sigma}\,\overline{g}^{\beta\rho}$,  
while the r.h.s. is the Landau-Lifschitz complex, where $t_{\rm LL}^{\alpha\beta}$ is the Landau-Lifschitz pseudotensor  
which represents, roughly to speak, the energy-momentum distribution of the gravitational fields. The field equations (\ref{Landau_Lifschitz_1})  
are manifestly Lorentz-covariant and constitute a set of ten coup\-led non-linear partial differential equations for
the ten components of the metric density $\overline{g}^{\alpha\beta}$.
Because of the identity $H^{\alpha\rho\beta\sigma}_{\;\;\;\;\;\;\;\;,\,\rho\sigma\beta} = 0$  
(implying energy-momentum conservation, cf. Eqs.~(6.7) - (6.8) in {\it Poisson,Will} (2014)) only six equations are independent, which  
determine the ten components of the metric density up to a passive transformation of coordinates which map the flat background manifold.  

Thus far, no specific choice of the coordinates of the flat background manifold has been imposed.    
For practical calculations in celestial mechanics, in the theory of light propagation, or in the theory of gravitational waves,  
it is, however, very useful to choose harmonic coordinates to cover the flat background space-time ${\cal M}_0$,   
which are introduced by the gauge condition
\begin{equation}
\overline{g}^{\alpha\beta}_{\;\;\;,\,\beta}\left(x\right) = 0 \quad\quad \Longrightarrow \quad\quad \square_g\,x^{\alpha} = 0 
\label{Landau_Lifschitz_3}
\end{equation}

\noindent
where the relation on the r.h.s. follows from the relation on the l.h.s. where $\square_g$ is the covariant d'Alembert operator which in harmonic coordinates  
reads $\square_g = g^{\rho\sigma}\,\nabla_{\rho}\,\nabla_{\sigma}$ and $\nabla_{\rho}$ denotes covariant derivative with respect to $x^{\rho}$. Harmonic  
coordinates are small deformations of the Minkowski coordinates, therefore it is useful to decompose the pulled-back metric density into the flat Minkowskian  
metric plus a small perturbation,  
\begin{equation}
\overline{g}^{\alpha\beta}\left(x\right) = \eta^{\alpha\beta} - \overline{h}^{\alpha\beta}\left(x\right) 
\label{Landau_Lifschitz_4}
\end{equation}

\noindent 
so that the gothic metric perturbation $\overline{h}^{\alpha\beta}$ propagates as dynamical field on the  
flat background space-time ${\cal M}_0$ (Section $7.1$ in {\it Carroll} (2013) and Section $6.2$ in {\it Poisson,Will} (2014)).  
By inserting (\ref{Landau_Lifschitz_3}) and (\ref{Landau_Lifschitz_4}) into (\ref{Landau_Lifschitz_1}) one obtains the Landau-Lifschitz field equations 
(also known as reduced field equations of gravity) in the following form (Eq.~(5.2b) in {\it Thorne} (1980))  
\begin{eqnarray}
\square\,\overline{h}^{\alpha\beta}\left(x\right) = - \frac{16\,\pi\,G}{c^4}\,\left(\tau^{\alpha \beta}\left(x\right) + t^{\alpha \beta}\left(x\right)\right)
\label{Field_Equations_10}
\end{eqnarray}

\noindent
where $\square = \eta^{\rho\sigma}\,\partial_{\rho} \partial_{\sigma}$ is the flat d'Alembert operator in terms of  
harmonic coordinates in the flat background space-time ${\cal M}_0$. The terms on the r.h.s. in (\ref{Field_Equations_10}) are given by
\begin{eqnarray}
	\tau^{\alpha \beta} = \left(- g\right)\,T^{\alpha \beta} \quad {\rm and} \quad  
t^{\alpha \beta} = \left(- g\right)\,t_{\rm LL}^{\alpha \beta} 
+ \frac{c^4}{16\,\pi\,G}
\left(\overline{h}^{\alpha\rho}_{\;\;\;,\;\sigma}\;\overline{h}^{\beta \sigma}_{\;\;\;,\;\rho} 
- \overline{h}^{\alpha\beta}_{\;\;\;,\;\rho\sigma}\;\overline{h}^{\rho\sigma}\right). 
\label{metric_40}
\end{eqnarray}

\noindent
The ten coupled non-linear partial differential equations (\ref{Field_Equations_10}) are exact field equations of gravity in the 
Landau-Lifschitz formulation in harmonic coordinates. Because of the gauge condition $\overline{h}^{\alpha\beta}_{\;\;\;,\,\beta}=0$,  
which follows from (\ref{Landau_Lifschitz_3}) and (\ref{Landau_Lifschitz_4}), only six equations are independent of each other.  

The harmonic gauge (\ref{Landau_Lifschitz_3}) does not uniquely select one harmonic coordinate system but a class of infinitely many harmonic systems,
because it allows for a residual gauge transformation between two arbitrary harmonic reference systems $\{x\}$ and $\{x^{\prime}\}$,
\begin{eqnarray}
        x^{\prime\,\alpha} = x^{\alpha} + \varphi^{\alpha}\left(x\right) 
\label{residual_gauge_transformation}
\end{eqnarray}

\noindent
if the gauge vector $\varphi^{\alpha}$ satisfies the homogeneous Laplace-Beltrami equation $\square_g\,\varphi^{\alpha} = 0$;  
Eq.~(\ref{residual_gauge_transformation}) has been elucidated by Fig.~$1$ in {\it Zschocke} (2019). The field equations (\ref{Field_Equations_10}) are 
invariant under the residual gauge transformation (\ref{residual_gauge_transformation}), which permits extensive simplifications of the form of 
the metric density. Moreover, the calculations of the MPM formalism are considerably simplified by assuming  
that $\{x\}$ are just Minkowskian (i.e. straight harmonic) coordinates, while $\{x^{\prime}\}$ are considered as curvilinear harmonic coordinates.

	\vspace*{0.7cm}
        \noindent {\large\bf 4. THE MULTIPOLAR POST-MINKOWSKIAN FORMALISM}
        \smallskip

The MPM approach has originally been introduced in {\it Thorne} (1980), while considerable extensions and 
important advancements have later been worked out in {\it Blanchet,Damour} (1986) and in a series of subsequent investigations. 
The MPM formalism is based on the post-Minkowski expansion of the field equations (\ref{Field_Equations_10}),  
\begin{eqnarray}
\overline{h}^{\alpha\beta} = \sum\limits_{n=1}^{\infty} G^n\,\overline{h}_{\left({\rm nPM}\right)}^{\alpha\beta} \quad {\rm and} \quad   
\tau^{\alpha\beta} = T^{\alpha\beta} + \sum\limits_{n=1}^{\infty} G^n\,\tau^{\alpha \beta}_{\left({\rm n PM}\right)} \quad {\rm and} \quad   
t^{\alpha\beta} = \sum\limits_{n=1}^{\infty} G^n\,t^{\alpha \beta}_{\left({\rm n PM}\right)}\;.     
\label{metric_53}
\end{eqnarray}

\noindent
Inserting (\ref{metric_53}) into (\ref{Field_Equations_10}) yields a hierarchy of field equations,  
\begin{eqnarray}
	&& \hspace{-0.5cm} \square\; \overline{h}_{\left({\rm 1PM}\right)}^{\alpha\beta}\left(x\right) = - \frac{16\,\pi}{c^4}\,T^{\alpha\beta}\left(x\right), 
\label{field_equation_1PM}
\\
\nonumber\\
&& \hspace{-0.5cm} \square\; \overline{h}_{\left({\rm 2PM}\right)}^{\alpha\beta}\left(x\right) = - \frac{16\,\pi}{c^4} 
\left(\tau_{\left({\rm 1PM}\right)}^{\alpha\beta}\left(x\right) + t_{\left({\rm 1PM}\right)}^{\alpha\beta}\left(x\right)\right),
\label{field_equation_2PM}
\\
\vdots 
\nonumber\\ 
&& \hspace{-0.5cm} \square\; \overline{h}_{\left({\rm nPM}\right)}^{\alpha\beta}\left(x\right) = - \frac{16\,\pi}{c^4}
\left(\tau_{\left({\rm (n-1) PM}\right)}^{\alpha\beta}\left(x\right) + t_{\left({\rm (n-1) PM}\right)}^{\alpha\beta}\left(x\right)\right).  
\label{field_equation_nPM}
\end{eqnarray}

\noindent
Each of the field equations (\ref{field_equation_1PM}) $\cdots$ (\ref{field_equation_nPM}) represents an equation in flat space-time. The MPM formalism is  
an approach for solving that hierarchy of field equations iteratively, starting  with the first iteration (\ref{field_equation_1PM}), where $T^{\alpha\beta}$  
is the energy-momentum tensor of matter in the approximation of special relativity. The general solution of the gothic metric in linear-order 
$\overline{h}_{\left({\rm 1PM}\right)}^{\alpha\beta}$ ({\it Thorne} (1980), {\it Blanchet,Damour} (1986), {\it Damour,Iyer} (1991))  
is inserted into the second iteration (\ref{field_equation_2PM}) which yields  
the gothic metric in post-linear order, $\overline{h}_{\left({\rm 2PM}\right)}^{\alpha\beta}$, and so on.  
Using this iterative approach, it has been demonstrated in {\it Blanchet,Damour} (1986) that the general solution of these field equations depends 
on six source-multipoles, $I_L,J_L,W_L,X_L,Y_L,Z_L$, which are integrals over the energy-momentum tensor of the compact source of matter  
(cf. Eqs.~(5.15) - (5.20) in {\it Blanchet} (1998)). Furthermore, using the residual gauge freedom (\ref{residual_gauge_transformation}), 
it has been demonstrated in {\it Blanchet,Damour} (1986) that the  
general solution of (\ref{field_equation_1PM}) $\cdots$ (\ref{field_equation_nPM}) can be written as follows,  
\begin{eqnarray}
        \overline{g}^{\alpha\beta}\left[I_L,J_L,W_L,X_L,Y_L,Z_L\right] &=& \eta^{\alpha\beta} 
        - \sum\limits_{n=1}^{\infty} G^n\,\overline{h}_{\left({\rm nPM}\right)}^{\alpha\beta\,{\rm can}}\left[M_L,S_L\right]
        + \; {\rm gauge}\;{\rm terms}   
\label{gothic_metric}
\end{eqnarray}

\noindent
which is valid in the exterior of the body.  
The canonical piece, $\overline{h}_{\left({\rm nPM}\right)}^{\alpha\beta\,{\rm can}}$,  
depends on two multipoles: mass-type multipole $M_L$ (accounts for shape, inner structure, and oscillations of the body)  
and current-type multipole $S_L$ (accounts for rotational motions and inner currents of the body), which are related to the source-multipoles  
via non-linear equations (Eqs.~(6.1a) and (6.1b) in {\it Blanchet} (1998)). 
All those terms in the metric density which depend on the gauge vector $\varphi^{\alpha}$ are called gauge terms and represent  
unphysical degrees of freedom because they have no impact on physical observables which are, by definition, coordinate-independent scalars 
({\it Bergmann} (1961)).  

The MPM formalism has been developed for understanding the generation of gravitational waves by an isolated source of matter, like  
binary black holes. Gravitational waves decouple from the source in the intermediate zone and they do finally propagate with the speed of light  
into the far wave-zone of the gravitational system. In the far wave-zone the gravitational fields have two degrees of freedom, where the   
transverse traceless (TT) gauge of the metric tensor becomes relevant because the TT terms in the metric tensor carry the physical information  
({\it Blanchet,Kopeikin,Sch\"afer} (2001)). In the far wave-zone, the TT projection of the metric density  
equals the TT projection of the metric tensor (cf. Eq.~(7.119) in {\it Carroll} (2013)),  
\begin{eqnarray}
	\overline{h}^{\rm TT}_{\alpha\beta} = h^{\rm TT}_{\alpha\beta} \quad \quad {\rm in}\;{\rm the}\;{\rm far - zone}\;.   
	\label{TT_metric_density}
\end{eqnarray}

\noindent
That is why there is no need to determine the metric tensor in the far wave-zone of the system. 
The gothic metric perturbation in TT gauge in terms of radiative moments $U_L$ and $V_L$, which are time-derivatives of source multipoles, 
is given by Eq.~(64) in {\it Blanchet,Kopeikin,Sch\"afer} (2001).

        \vspace*{0.7cm}
        \noindent {\large\bf 5. THE METRIC TENSOR}
        \smallskip

For determining light trajectories in the near-zone of the solar system one needs the metric tensor of solar system bodies.  
While in principle one might use the TT gauge, one should, however, not expect much simplification,  
because such a nice relation like (\ref{TT_metric_density}) does not exist,  
\begin{eqnarray}
	\overline{h}^{\rm TT}_{\alpha\beta} \neq h^{\rm TT}_{\alpha\beta} \quad \quad {\rm in}\;{\rm the}\;{\rm near - zone}\;.   
        \label{TT_metric_tensor}
\end{eqnarray}

\noindent 
Thus, relativistic astrometry necessarily requires the determination of the metric tensor in the near-zone of the gravitational system. 
The metric density and the metric tensor contain the same physical information 
about the gravitational system, because they are related to each other reversibly unique by Eqs.~(\ref{Metric_Density}).  
Using these relations, it has been shown  
in {\it Zschocke} (2019) that the general form of the metric tensor in the exterior of a solar system body is given by  
\begin{eqnarray}
        g_{\alpha\beta}\left[I_L,J_L,W_L,X_L,Y_L,Z_L\right] &=& \eta_{\alpha\beta}
	+ \sum\limits_{n=1}^{\infty} G^n\,h^{\left({\rm nPM}\right)}_{\alpha\beta\,{\rm can}}\left[M_L,S_L\right]
        + \; {\rm gauge}\;{\rm terms} 
\label{metric}
\end{eqnarray}

\noindent
where the canonical piece, $h^{\left({\rm nPM}\right)}_{\alpha\beta\,{\rm can}}$, depends only on two multipoles $M_L$ and $S_L$. 
The linear term and the post-linear term of the metric perturbation, $h^{\left({\rm 1PM}\right)}_{\alpha\beta\,{\rm can}}$ and  
$h^{\left({\rm 2PM}\right)}_{\alpha\beta\,{\rm can}}$, respectively, are explicitly given by Eqs.~(109) - (111) and (115) - (117) in {\it Zschocke} (2019).  
The gauge terms depend on the gauge vector $\varphi^{\alpha}$ and have no impact on physical observables.  

	\vspace*{0.7cm}
	\noindent {\large\bf 7. CONCLUSION}
	\smallskip
	
Future astrometry at the sub-micro-arcsecond and nano-arcsecond level of accuracy in astrometric measurements requires considerable progress in modeling the  
trajectory of light signals through the curved space-time of the solar system. Such a precise determination of light trajectories implies the knowledge of the   
metric tensor $g_{\alpha\beta}$ of solar system bodies in the post-linear approximation.  
The Multipolar Post-Minkowskian formalism represents a framework for determining the metric density $\overline{g}^{\alpha\beta}$
in the exterior of a massive body having arbitrary shape and inner structure, oscillations and rotational motions. The knowledge of
the metric density allows to deduce the metric tensor $g_{\alpha\beta}$. Some aspects of that approach have been considered 
which are relevant for future investigations in the theory of light propagation and relativistic astrometry.

	\vspace*{0.7cm}
	\noindent{\large\bf 8. REFERENCES}
	{
	
	\leftskip=5mm
	\parindent=-5mm
	\smallskip

        Bergmann, P.G., 1961, "Observables in General Relativity",  
	Rev. Mod. Phys. 33, pp. 510-514.  

        Blanchet, L., 1998, "On the multipole expansion of the gravitational field",
        Class. Quantum Grav., 15, pp. 1971-1999.

        Blanchet, L., Damour, T., 1986,
        "Radiative gravitational fields in general relativity: I. General structure of the field outside the source",
        Phil. Trans. R. Soc. London A, 320, pp. 379-430.
	
        Blanchet, L., Kopeikin, S.A., Sch\"afer, G., 2001, "Gravitational radiation theory and light propagation",
        Lecture Notes in Physics, 562, pp. 141-166.

        Brian Pitts, J., Schieve W.C., 2018, "Null Cones in Lorentz-Covariant General Relativity",  
	in gr-qc/0111004.  

        Carroll, S., 2013, "Spacetime and Geometry: An Introduction to General Relativity", First Edition,  
        Pearson New International, Edinburgh Gate, UK.

        Damour, T., Iyer, B.R., 1991, "Multipole analysis for electromagnetism and linearized gravity with irreducible Cartesian tensors", 
	Phys. Rev. D, 43, pp. 3259-3272.  

        Earman, J., Norton, J., 1987,  
	"What price spacetime substantivalism? The hole story",  
	The British Journal for the Philosophy of Science 38 (4), pp. 515-525.  

        Feynman, R.P., 1995,  
        "Lectures on Gravitation", First Edition,  
        Addison-Wesley, Boston, U.S.  

	Gaul, M., Rovelli, C., 2000,  
        "Loop Quantum Gravity and the Meaning of Diffeomorphism Invariance", 
	Lect. Notes Phys. 541, pp. 277-324.  

        Gupta, S.N., 1954, "Gravitation and Electromagnetism",  
	Phys. Rev., 96, pp. 1683-1685.  

        Hawking, S.W., Ellis, G.F.R., 1974, "The large scale structure of space-time", First Edition,  
        Cambridge University Press, New York, U.S.

        Landau, L.D., Lifschitz, E.M., 1971, "The Classical Theory of Fields",
        Third English Edition, Course of Theoretical Physics, Volume 2,
        Pergamon Press, Oxford, UK.

        Lusanna, L., Pauri, M., 2006,
        "The Physical Role of Gravitational and Gauge Degrees of Freedom in General Relativity - I,II" 
        Gen. Rel. Grav. 38, pp. 187-227, pp. 229-267.  

	Misner, C.W., Thorne, K.S., Wheeler, J.A., 1973, "Gravitation", First Edition,  
	W.H. Freeman, New York, U.S. 

        Petrov, A.N., Kopeikin, S.M., Lompay, R.R., Tekin, B., 2017,
        "Metric Theories of Gravity: Perturbations and Conservation Laws", First Edition,
        De Gruyter, Boston, U.S.

        Poisson, E., Will, C.M., 2014, 
        "Gravity: Newtonian, Post-Newtonian, Relativistic", First Edition, Cambridge University Press, UK. 

        Thorne, K.S., 1980, "Multipole expansions of gravitational radiation",  
        Rev. Mod. Phys., 52, pp. 299-339.

        Zschocke, S., 2019, "Post-linear metric of a compact source of matter",  
        Phys. Rev. D, 100, 084005, pp. 1-32.
  
        }

       \end{document}